\documentstyle[epsf,psfig]{elsart}
\begin{document}
\begin{frontmatter}
\title{Formation of binary correlations in strongly coupled plasmas}
\author{K. Morawetz}
\address{
Fachbereich Physik, Universit\"at Rostock,
18051 Rostock, Germany}
\author{V\'aclav \v Spi\v cka and Pavel Lipavsk\'y}
\address{Institute of Physics, Academy of Sciences, Cukrovarnick\'a 10,
16200 Praha 6, Czech Republic}
\date{\today}
pacs: 05.60.+w,82.20.Mj,52.25.Dg,82.20.Rp

\begin{abstract}
Employing quantum kinetic equations we study the formation of binary correlations in plasma at short time scales. It is shown that this formation is much
faster than dissipation due to collisions, in hot (dense) plasma the
correlations form on the timescale of inverse plasma frequency (Fermi
energy). This hierarchy of characteristic times is used to derive
analytical formulae for time dependency of the potential energy of
binary interactions which measures the extent of correlations. We discuss the 
dynamical formation of screening and compare with the static screened 
result. Comparisons are made with molecular dynamic simulations. In the low 
temperature
limit we find an analytical expression for the formation of correlation which 
is general for any binary interaction. It can be applied in nuclear situations 
as well as dense metals.
\end{abstract}
\end{frontmatter}

Recent lasers allow one to create a high density plasma within few
femto seconds and observe its time evolution on a comparable scale
\cite{HJ96,THWS96}. Naturally, this plasma is highly excited at the
beginning and relaxes towards equilibrium by various mechanisms that
might be dominant at some stage and sub-dominant in another one. The
best known regimes are the fast local equilibration of electron and
hole distributions due to binary collisions, and the slow global
relaxation via diffusion, recombination, dissipation of energy into
the host crystal, etc. There are, however, even faster processes than
the local equilibration. These processes dominate during the very
first stage of the relaxation, the so called transient regime. In this
Letter we discuss the transient regime in terms of the energy balance.

The transient regime has been already discussed from many different
angles and for various systems (classical -- quantum, non-degenerate
-- degenerate). Let us briefly review some of these approaches. We
will show that they are equivalent, at least with respect to the
energy balance.

Conceptually the simplest is the molecular dynamics. One takes $N$
particles, distributes them randomly into a box and let them
classically move under Coulomb forces due to their own charges.
Those particles which are very close will be expelled from each
other. Their first movement thus forms correlations which lower
the Coulomb energy $V_{\rm C}=e^2/r$. This build up of screening
stops when the effective Debye potential
$V_{\rm D}=e^2 {\rm e}^{-\kappa r} /r$ is reached. An important
question is whether the long-range or the short-range charge
fluctuations dominate in this process. In the former case, the
characteristic time of the transient period should be the inverse
plasma frequency $\tau_c=1/\omega_p$. In the latter case we do not know.

If one ``measures'' all states of the systems by their distance from
equilibrium, instead of {\em formation} of correlations one has to
talk about their {\em decay}. Our presumption that the transient
period is appreciably shorter than the local relaxation is thus just
Bogolyubov's principle of decay of correlations. The decay of
correlations is linked with an alternative approach to the transient
period, the formation of quasi-particles which has been numerically
studied within Green's functions \cite{BKSBKK96}. We note that very
similar transient behavior has been observed for the nuclear matter
\cite{D841,K95}, i.e., the formation/decay of correlation is a rather
general phenomenon.

The first concept is based on the two-particle space correlations while
the second one on the single-particle excitations. The quantity which
allows us to follow both pictures in a unified manner is the energy of the 
system. It is composed from the kinetic energy $\left\langle{k^2\over
2m}\right\rangle$ and the correlation energy $E_{\rm corr}=\frac 1 2 \langle
V_{\rm D}\rangle -\langle V_{\rm C}\rangle$, where $\langle V_{\rm
C}\rangle$ subtracts the background. In lowest order interaction and classical limit the correlation energy takes the Debye H\"uckel form
\begin{equation}
E_{\rm corr}=\frac 1 2 \lim\limits_{r \to 0} \left [ V_D(r)-V_C(r) \right ]=-{\kappa e^2 \over 2}.\label{deb}
\end{equation}
Of course, the total energy
conserves,
\begin{equation}
E_{\rm corr}=\left\langle\frac{k^2}{2m}\right\rangle_0-
\left\langle\frac{k^2}{2m}\right\rangle,
\label{cons}
\end{equation}
where $\left\langle{k^2\over 2m}\right\rangle_0$ is the initial value
of the kinetic energy. We will monitor the time dependency of the
transfer of the correlation energy into the kinetic one.

It is more convenient to calculate the kinetic energy than the
correlation energy because the kinetic one is a single-particle
observable. To this end we can use the kinetic equation, of course,
an equation which leads to the total energy conservation (\ref{cons}). It is
immediately obvious that the ordinary Boltzmann equation cannot be
appropriate for this purpose because the kinetic energy is an invariant
of its collision integral and thus constant in time. We have to consider
non-Markovian kinetic equations of Levinson type \cite{HJ96}
\begin{eqnarray}
\frac{\partial}{\partial t}f_a(t)&=&\frac{2}{\hbar^2}\sum\limits_b
\int\frac{dpdq}{(2\pi\hbar)^6}V_{\rm D}^2(q)
\int\limits_0^t d\bar t\,
\exp\left\{-{t-\bar t\over\tau}\right\}\,
{\rm cos}\left\{\frac{1}{\hbar}(t-\bar t)\Delta_E\right\}
\nonumber\\
&&\times\left\{\bar f'_a\bar f'_b(1\!-\!\bar f_a)(1\!-\!\bar f_b)-
\bar f_a\bar f_b(1\!-\!\bar f'_a)(1\!-\!\bar f'_b)\right\},
\label{kinetic}
\end{eqnarray}
where $\Delta_E={k^2\over 2m_a}+{p^2\over 2m_b}-{(k-q)^2\over 2m_a}-
{(p+q)^2\over 2m_b}$ denotes the energy difference between initial and final
states. The retardation of distributions, $\bar f_a(k,\bar t)$,
$\bar f'_a(k-q,\bar t)$ etc., is balanced by the lifetime $\tau$. The
total energy conservation (\ref{cons}) for Levinson's equation has been
proved in \cite{M94}.

The full solution of Levinson's equation on the long time scale is a
hard problem, however, its solution in the short-time region $t\ll\tau$
can be written down analytically. In this time domain we can neglect the
time evolution of distributions, $\bar f_a(\bar t)=f_a(0)$, and the life-time
factor, $\exp\left\{-{t-\bar t\over\tau}\right\}=1$. The eq. (\ref{kinetic}) can be integrated with respect to the time and the internal time integral can be done. The resulting equation for $f(t)$ represents
the deviation of Wigner's distribution from its initial value, $f_a(t)=
f_a(0)+\delta f_a(t)$, and reads
\begin{eqnarray}\label{short1}
\delta f_a(t)&=&2\sum\limits_b
\int\frac{dpdq}{(2\pi\hbar)^6}V^2_{\rm D}(q)
{1-\cos\left\{{1\over\hbar}t\Delta_E\right\}\over\Delta_E^2}
\nonumber\\
&&\times\left\{f'_a f'_b(1\!-\!f_a)(1\!-\!f_b)-f_a
f_b(1\!-\!f'_a)(1\!-\!f'_b)\right\}.
\end{eqnarray}
This formula shows how the two-particle and the single-particle concept
of the transient behavior meet in the kinetic equation. The right hand
side describes how two particles correlate their motion to avoid the
strong interaction regions. Since the process
is very fast, the on-shell contribution to $\delta f_a$, proportional
to $t/\tau$, can be neglected in the assumed time domain and
the $\delta f$ has the pure off-shell character as can be seen from
the off-shell factor $\Delta_E^{-2}\left(1-
\cos\left\{{1\over\hbar}t\Delta_E\right\}\right)$. 
The off-shell character of mutual
two-particle correlation is thus reflected in the single particle
Wigner's distribution.

The
very fast formation of the off-shell contribution to Wigner's
distribution has been found in numerical treatments of Green's
functions \cite{D841,K95}. Once formed, the off-shell contributions
change in time with the characteristic time $\tau$, i.e., following
the relaxation (on-shell) processes in the system. Accordingly, the
formation of the off-shell contribution signals that the system has
reached the state the evolution of which can be described by the
Boltzmann equation, i.e., the transient period has been accomplished.

From Wigner's distribution one can readily evaluate the increase of
kinetic energy,
\begin{equation}
\left\langle{k^2\over 2m}\right\rangle-
\left\langle{k^2\over 2m}\right\rangle_0
=\sum_a\int{dk\over(2\pi\hbar)^3}{k^2\over 2m_a}\delta f_a.
\label{kine}
\end{equation}
After substitution for $\delta f_a$ from (\ref{short1}) we symmetrize
in $k$ and $p$ and anti-symmetrize in the initial and final states which
yields the correlation energy (\ref{cons}) as
\begin{eqnarray}
E^{\rm static}_{\rm corr}(t)&=&-\sum_{ab}\int\frac{dkdpdq}{(2\pi\hbar)^9}V^2_{\rm D}(q)
\frac{1-\cos\left\{{1\over\hbar}t\Delta_E\right\}}{\Delta_E}
f'_a f'_b(1-f_a)(1-f_b).
\label{energ1}
\end{eqnarray}
This expression holds for general distributions $f_a$. 

Of course, starting with a sudden switching approximation we have Coulomb interaction and during the first transient time period the screening is formed. This can be described by the non-Markovian Lenard - Balescu equation \cite{Moa93} instead of the static screened equation (\ref{kinetic}). With the same discussion as above we end up instead of (\ref{energ1}) with the dynamical expression of the correlation energy
\begin{eqnarray}
E^{\rm dynam}_{\rm corr}(t)&=&-\sum_{ab}\int\frac{dkdpdq}{(2\pi\hbar)^9}{V^2_{\rm C}(q)\over |\epsilon(q,{(p+q)^2 \over 2 m_b}-{p^2 \over 2 m_b}) |^2}
\left ({(k-q)^2 \over 2 m_a}-{k^2 \over 2 m_a}\right )
\nonumber\\
&\times&
\frac{1-\cos \left \{ {1\over\hbar}t\Delta_E \right \}}{\Delta_E^2} f'_a f'_b (1-f_a)(1-f_b).\nonumber\\
&&
\label{energ2}
\end{eqnarray}
One sees that the bare Coulomb interaction $V_C$ is renormalized by the dielectric function
\begin{equation}
\epsilon(q,\hbar \omega)=1-\sum\limits_b V_C(q)\int {d p\over (2 \pi \hbar)^3} {f_b(p+\frac 1 2 q)-f_b(p-\frac 1 2 q)\over {p q\over m_b} -\hbar \omega+i\eta}.
\end{equation}
All internal time integrals are bound to the time dependence of $f(t)$ and in the spirit of the above discussion can be carried out.

To demonstrate its
results and limitations, we discuss (\ref{energ1}) and (\ref{energ2}) for 
special cases
that allow for analytical treatment. To this end we use equilibrium
initial distributions. 
As the first test, let us evaluate the correlation energy in
equilibrium which is approached for large times $t\to\infty$. The
off-shell factor ${1\over\Delta_E}\left(1-\cos\left\{{1\over\hbar}t
\Delta_E\right\}\right)$ then turns into the principle value
$\frac{\wp}{\Delta E}$. The equilibrium distributions $f_a$ are natural
for this case.

At the high temperature limit, where the distributions are
non-degenerate
\begin{eqnarray}
f_a=n_a\hbar^3\left({2\pi\over m_aT}\right)^{{3\over 2}}
\exp\left\{-{k^2\over 2m_aT}\right\},
\label{dist}
\end{eqnarray}
one can evaluate (\ref{energ1}) and (\ref{energ2}) 
for the mixture of particles.
We assume the plasma consisting of two different types of particles
$a,b$ with different masses $m_a,m_b$. Performing a
series of integrals we obtain the correlation energies 
\begin{eqnarray}
E^{\rm static}_{\rm corr}(\infty)&=&-\pi\sum\limits_{ab}\left({4 m_am_b\over(m_a+m_b)^2}\right)^2
{e_a^2e_b^2n_an_b\over\kappa T}
\left[1-\sqrt{\pi}b\exp(b^2){\rm erfc}(b)\right],
\nonumber\\
E^{\rm dynam}_{\rm corr}(\infty)&=&- \pi^{3/2}\sum\limits_{ab}\left({4 m_am_b\over(m_a+m_b)^2}\right)^2
{2 e_a^2e_b^2n_an_b\over b\ \kappa T}
\left[1-\exp(b^2){\rm erfc}(b)\right].
\nonumber\\
&&
\label{vt}
\end{eqnarray}
The parameter $b^2=(\hbar\kappa)^2{m_a+m_b\over 8m_am_b T}$ controls
quantum corrections. 
Formula (\ref{vt}) is the correlation energy in the
second Born approximation of statically screened Debye potential as well as dynamically screened one. 
The latter corresponds to known Montroll result in
plasma physics \cite{kker86,RSWK95}.

For identical particles in the classical limit $b\to 0$, we see
that the static approximation (\ref{energ1}) or (\ref{vt}) underestimates the 
known value of the correlation energy \cite{kker86,RSWK95} while the dynamical result (\ref{energ2}) or (\ref{vt}) agrees with this Ward result
\begin{eqnarray}
E_{\rm corr}^{\rm static}(\infty)&=&-{e^2 \kappa \over 4
} (1-\sqrt{\pi} \b \ {\rm erfc} (b))=-{1\over 4}e^2n\kappa+o(b),
\nonumber\\
E_{\rm corr}^{\rm dynam}(\infty)&=&-{e^2 \kappa \over 2
}{\sqrt{\pi} \over b} (1-{\rm e}^{b^2} {\rm erfc}
b)=-{1\over 2}e^2n\kappa+o(b).
\label{class}
\end{eqnarray}
One can see that in the classical limit the static result is just one half of the correct Debye-H\"uckel one (\ref{deb}). The dynamical result yields the correct correlation energy. This difference can be understood in analogy to the field energy of a dipole in an external electric field. If the dipole is already present but has just to be ordered, we obtain half of the correlation energy we would have if the dipole is formed itself by the field. In our case the static result assumes that we have a Debye screening from the beginning before the interaction is switched on. The dynamical result counts properly for the fact that the screening has to be formed itself which results into twice the correlation energy.

\begin{figure}
\parbox[t]{6.5cm}{
\psfig{figure=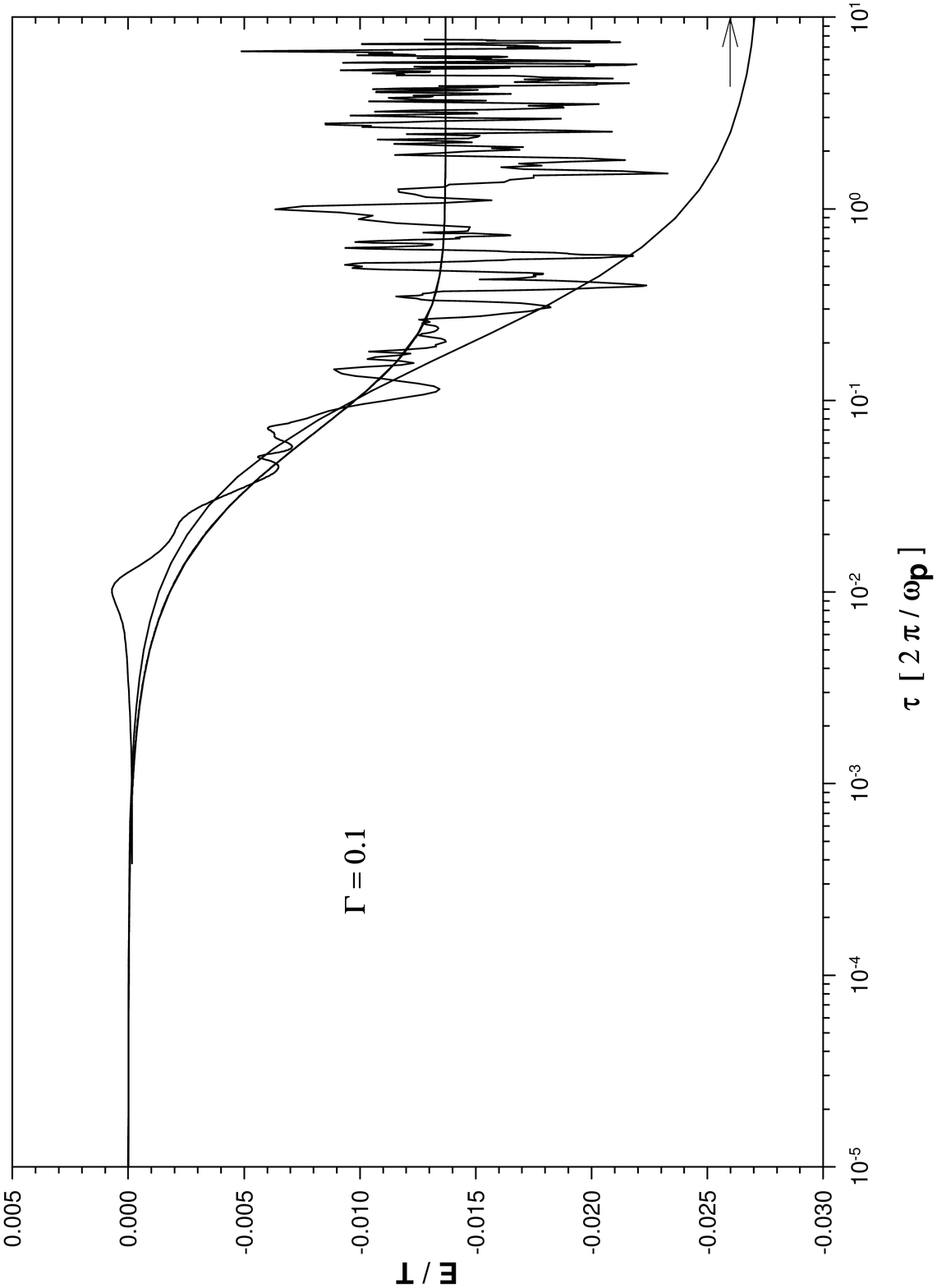,width=6.5cm,height=6cm,angle=-90}}
\hspace{1ex}
\parbox[t]{6.5cm}{
\psfig{figure=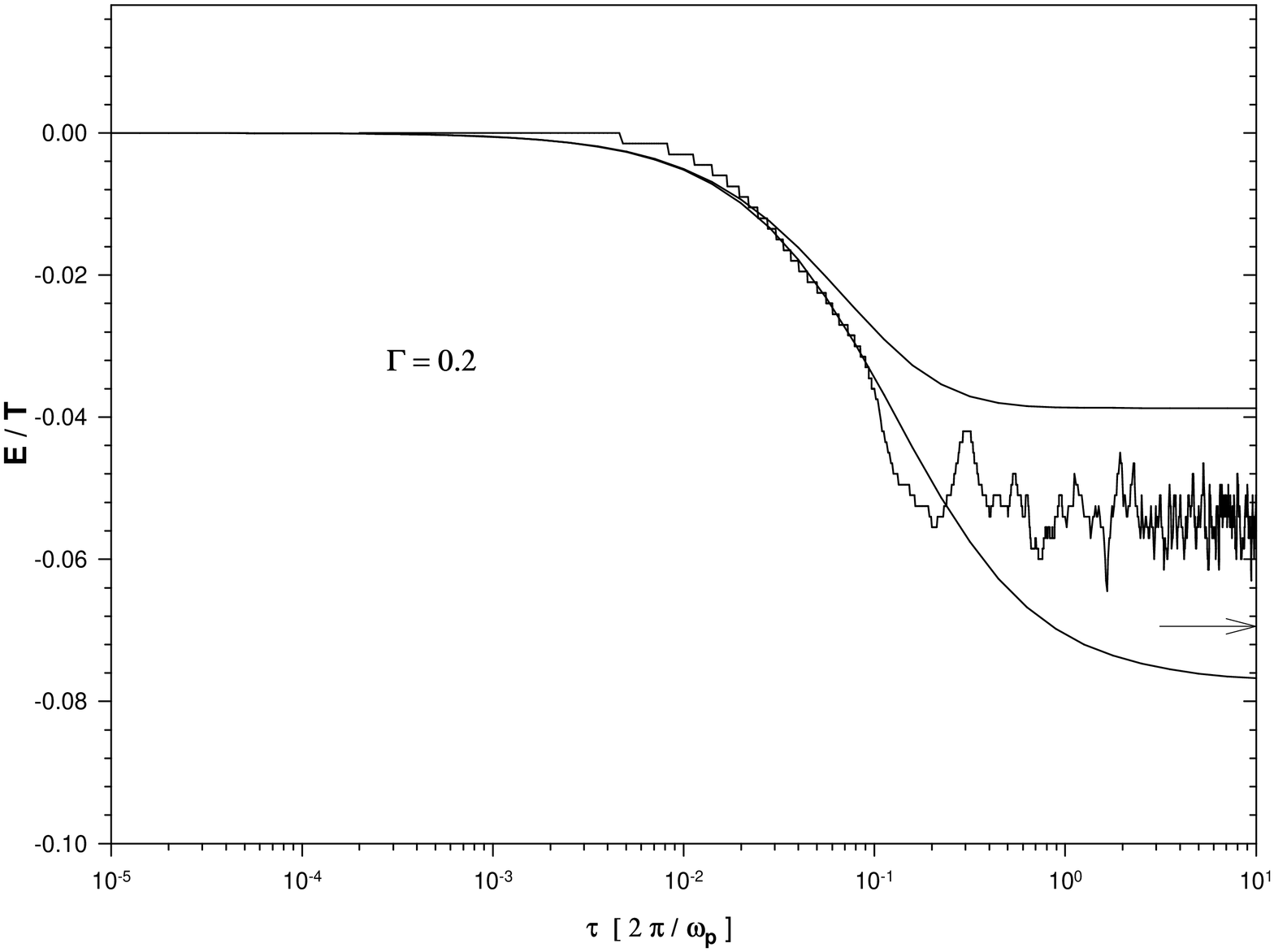,width=6.5cm,height=6cm,angle=-90}}
%\vspace{-4ex}
\caption{\label{1}The formation of correlation energy due to molecular
dynamic simulations \protect\cite{ZTRa95} together with the analytical 
result of
(\protect\ref{v2}) for a plasma parameter $\Gamma=0.1$ (left) and $\Gamma=0.2$ (right). The upper curve is the static and the lower the dynamical calculation. The latter one approaches the Debye-H\"uckel result. The exact equilibrium correlation energy of MC simulations \protect\cite{I94} are indicated by the arrow.}
\end{figure}

\begin{figure}
\parbox[t]{6.5cm}{
\psfig{figure=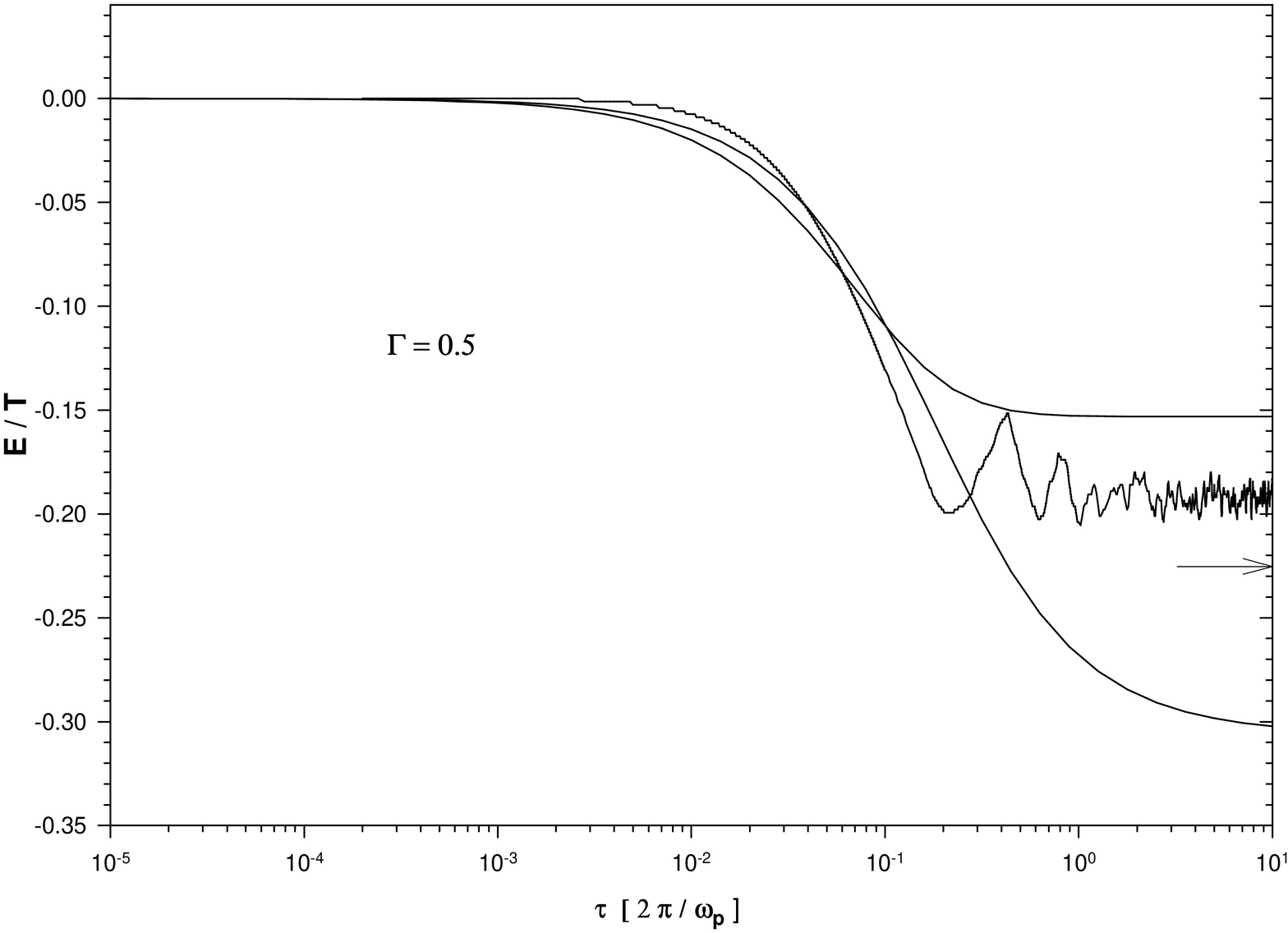,width=6.5cm,height=6cm,angle=-90}}
\hspace{1ex}
\parbox[t]{6.5cm}{
\psfig{figure=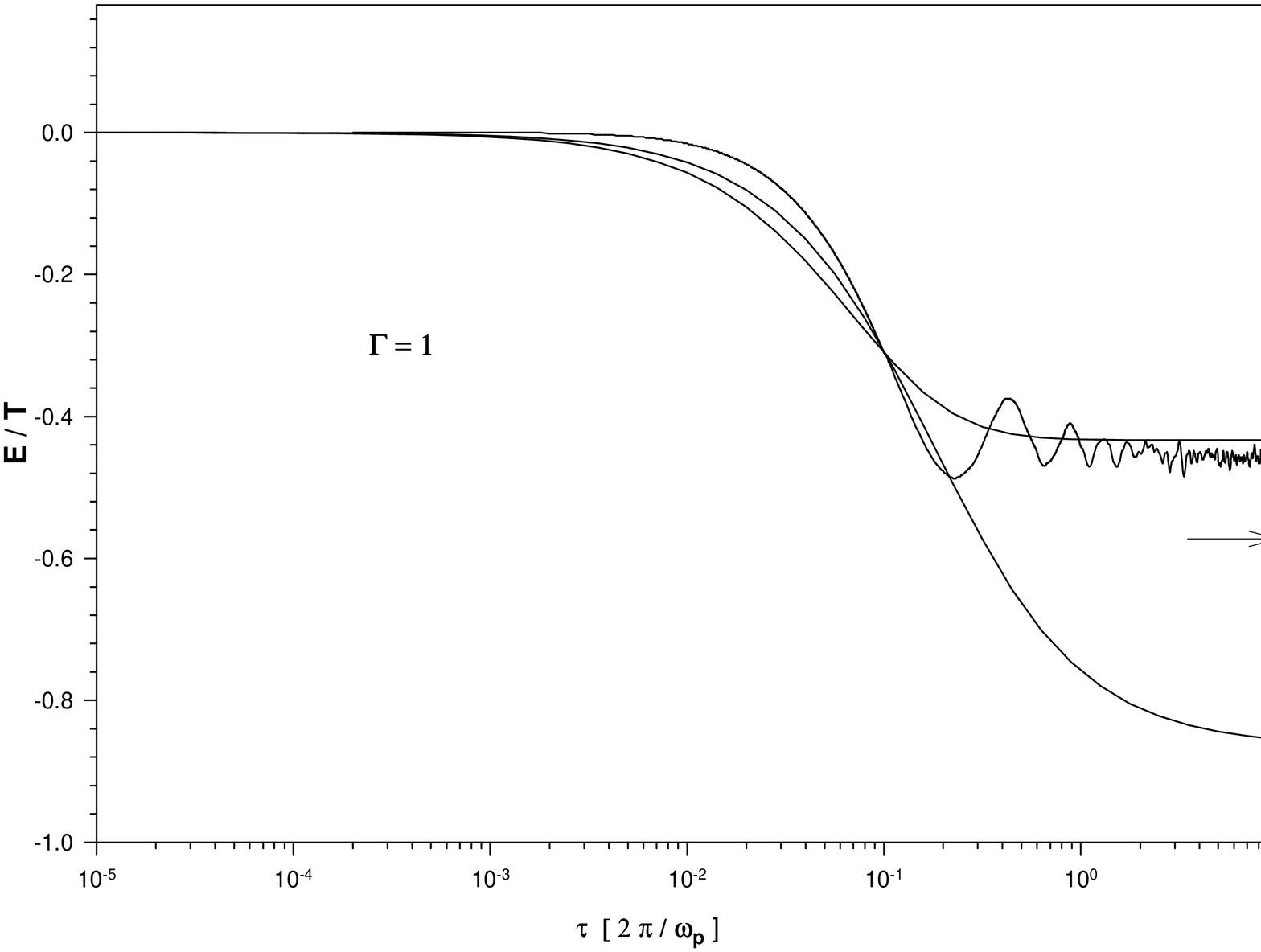,width=6.5cm,height=6cm,angle=-90}}
\caption{\label{2}The formation of correlation energy due to molecular
dynamic simulations \protect\cite{ZTRa95} together with the result of
(\protect\ref{v2}) for a plasma parameter $\Gamma=0.5$ (left) and $\Gamma=1$ (right). The upper curve is the static and the lower the dynamical calculation. The latter one approaches the Debye - H\"uckel result. The simulation approaches the exact correlation energy \protect\cite{I94} with higher densities indicated by the arrow.}
\end{figure}

In order to compare the time dependency of the correlation energy from
(\ref{energ1}) with molecular dynamical simulations \cite{ZTRa95}, we
assume a one component plasma which possesses a Maxwellian velocity
distribution (\ref{dist}) during this formation time. From
(\ref{energ1}) and (\ref{energ2}) we find
\begin{eqnarray}
{\partial \over \partial t} {E_{\rm corr}^{\rm static}(t) \over n}&=& -{e^2 \kappa T\over 2 \hbar}{\rm Im}
\left [(1+2 z^2 ) {\rm e}^{z^2} (1- {\rm erf} (z)) -{2 z \over \sqrt{\pi}} \right ], \nonumber\\
{\partial \over \partial t} {E_{\rm corr}^{\rm dynam}(t) \over n}&=& 
-{e^2 \kappa T \over  \hbar}{\rm Im}
\left [{\rm e}^{z_1^2} (1- {\rm erf} (z_1)) \right ] \nonumber\\&&
\label{v1}
\end{eqnarray}
where we used $z =\omega_p \sqrt{t^2 - i t {\hbar \over T}}$ and $z_1 =\omega_p \sqrt{2 t^2 - i t {\hbar \over T}}$.
This is the analytical quantum result of the time derivative of the formation of correlation for statically as well as dynamically screened potentials. For the classical limit $\hbar \to 0$ it is easy to integrate expression (\ref{v1}) with respect to times and arrive at
\begin{eqnarray}
E_{\rm corr}^{\rm static}(t)&=&-{1\over 4}e^2n\kappa
\Biggl\{1+{2\omega_p t\over\sqrt{\pi}}
 -\left(1+2\omega_p^2t^2\right)\exp\left(\omega_p^2t^2\right)
\left[1-{\rm erf}(\omega_p t)\right]\Biggr\},\nonumber\\
E_{\rm corr}^{\rm dynam}(t)&=&-{1\over 2}e^2n\kappa
\Biggl\{1-\exp\left({\omega_p^2 \over 2 } t^2\right)
\left[1-{\rm erf}({\omega_p \over \sqrt{2}} t)\right]\Biggr\}.
\nonumber\\
&&
\label{v2}
\end{eqnarray}
In Figs.~\ref{1} and \ref{2}, this formulae are compared with molecular
dynamic simulations \cite{ZTRa95} for different values of the plasma parameter
$\Gamma$. The parameter $\Gamma={e^2\over a_eT}$, where
$a_e=({3\over 4\pi n})^{1/3}$ is the inter-particle distance or
Wigner-Seitz radius, measures the strength of the Coulomb coupling.
Ideal plasma are found for $\Gamma\ll1$. In this region the static formula
(\ref{v2}) well follows the major trend of the numerical result, see
Fig.~\ref{1}. The agreement is in fact surprising, because we saw that the static result underestimates the dynamical long time result of Debye- H\"uckel (\ref{deb}) 
${\kappa e^2 \over 2}=\sqrt{3} /2 \Gamma^{3/2}$ by a factor of two.
%\vspace{-4ex}

The explanation for this fact is that we can prepare the initial configuration within our kinetic theory such that sudden switching of interaction is fulfilled. However, in the simulation experiment we have initial correlations which are due to the setup within quasiperiodic boundary condition and Ewald summations. This obviously results into an effective statically screened Debye potential, or at least the simulation results allow for this interpretation.

If we go to higher densities of $\gamma=0.2$ in figure \ref{1} (right) we see that the Debye H\"uckel result is far off the correct equilibrium correlation energy. Nevertheless the time scale is still appropriate described by the dynamical result.
At still higher densities, like $\Gamma=0.5$ and $\Gamma=1$, see Fig.~\ref{2}, non-ideal effects become important and
the formation time is underestimated within (\ref{v2}).
Of course, for higher plasma parameter the Born approximation
fails. Nevertheless, formula (\ref{v2}) can still almost reproduce the
formation time but slightly shorter than compared with
the simulation. This is due to non-ideality which was found to be an
expression of memory effects \cite{MWR93} and leads to a later
relaxation. For strongly coupled plasma the Born approximation fails, of course,
to reproduce the correct equilibrium value, but reproduces the formation
time fairly good. The equilibrium value, of course, differs and takes smaller values than the Debye - H\"uckel result \cite{I94,DSC96}. This regime is clearly out of scope of our theory.

The characteristic time of formation of correlations at high temperature
limit is given by the time where (\ref{v2}) shows a saturation. This is
reached at about a time of inverse plasma frequency
\begin{eqnarray}
\tau_c\approx{1\over\omega_p}={\sqrt{2}\over v_{\rm th}\kappa}.
\end{eqnarray}
The inverse plasma frequency indicates that the dominant role is played by long
range fluctuation. On the other hand, we also see that the correlation time is found to be
given by the time a particle needs to travel through the range of the
potential with a thermal velocity $v_{\rm th}$. This confirms the
numerical finding of \cite{BKSBKK96} that the correlation or memory
time is proportional to the range of interaction, i.e. rather of short range character.

In the low temperature region, i.e., in a highly degenerate system
$\mu\gg T$, one finds a different picture. From (\ref{energ1}) follows
\cite{MK97}
\begin{eqnarray}
&&E^{low}_{\rm corr}(t)-E_{\rm corr}^{low}(0)= E_{\rm corr}^{\rm low}
{1-\frac 1 x \sin(x)
+\left ({2 \mu \over \pi T}\right )^2 \left (\frac 1 3 + \left [\frac 1 x \sin(x)\right ]''\right )
\over
1 +  \frac 1 3 ({2 \mu \over \pi T})^2}
\label{c1b}
\end{eqnarray}
with $x={4 \mu \over \hbar} t$ and the equilibrium correlation energy
\begin{eqnarray}
E_{\rm corr}&=&\frac{\mu e^2}{12 \kappa^3 \pi^2} \left (T^2+ \frac 1 3 \left ({2 \mu \over \pi}\right 
)^2\right ) \left ({m \over \hbar^2}\right )^4  \left (
\arctan{\frac {1}{ b_l}}+{b_l \over 1+b_l^2} \right )\nonumber\\
&&
\label{equil}
\end{eqnarray}
where the abbreviation is $b_l={\hbar\kappa\over 2p_F}$. We like to point out that the formula (\ref{c1b}) is
generally valid for any binary interaction and
applicable in dense metal as well as nuclear situations. The only
difference lies in the actual value of the equilibrium correlation energy which is dependent on the potantial, of course. Indeed, (\ref{c1b}) follows from the standard procedure 
of separating angular from energy integrals for low temperatures. The time dependence is carried exclusively by the energy integrals while the angular averaged interaction results in its factor.

Unlike in the classical case, the equilibrium limit of the degenerate
case (\ref{c1b}) is not reached monotonously but with oscillation that
are damped with power law $t^{-1}$ in time. In other words,
the correlation energy is rapidly built up and then oscillates around
the equilibrium value (\ref{equil}). We can define the build up time
$\tau_c$ as the time where the correlation energy reaches its first
maximum,
\begin{eqnarray}
\tau_c=1.05{\hbar\over\mu}
\label{exact}
\end{eqnarray}
with  $\mu$ the Fermi energy.
Note that $\tau_c$ is in agreement with the quasiparticle formation
time known as Landau's criterion. 
Indeed, as argued above, the
quasiparticle formation and the build up of correlations are two
alternative views of the same phenomenon.

The formation of binary correlations is very fast on the time scale of
dissipative process. With respect to dissipative regimes, the binary
correlations can be treated as instant functionals of the
single-particle distribution and thus included into the Boltzmann
equation via various renormalizations of its ingredients, would it be
the screened Coulomb potential in the scattering rate or the
quasiparticle corrections. Under extremely fast external perturbations,
like the massive femto second laser pulses, the dynamics of binary
correlations will hopefully become experimentally accessible. Even if
related measurement will not reveal any unexpected features, the
experimental justification of basic concepts of the non-equilibrium
many-body physics is very desirable. The theoretical support to such
experiments is mostly based on the non-equilibrium Green's functions or
the molecular dynamics which both demand expensive numerical treatments.
Unlike these two approaches, the presented theory fails for special
systems where the characteristic time of formation of correlations
becomes longer or comparable with other time scales. For normal systems,
it provides a simple tool for theoretical predictions.

We are grateful to G. Zwicknagel who was so kind as to provide the data
of simulations. Stimulating discussion with G. R{\"o}pke is
acknowledged. This project was supported by the BMBF
(Germany) under contract Nr. 06R0884, the Max-Planck Society, 
Grant Agency of Czech
Republic under contracts Nos. 202960098 and 202960021  and the EC Human Capital and
Mobility Programme.

%\bibliography{kmsr,kmsr1,kmsr2,kmsr3,kmsr4,kmsr5,kmsr6,kmsr7,delay1,spin}
%\bibliographystyle{prsty}

\end{document}